\documentclass[]{spie}  

\usepackage{amsmath,amsfonts,amssymb}
\usepackage{graphicx}
\usepackage[colorlinks=true, allcolors=blue]{hyperref}

\title{A software toolkit to simulate activation background for high energy detectors onboard satellites }

\author[a,b]{G\'abor Galg\'oczi}
\author[a,c,d]{Jakub \v{R}\'{i}pa}
\author[e,f]{Giuseppe Dilillo}
\author[a]{Masanori Ohno}
\author[g,h]{Riccardo Campana}
\author[c,a,d]{Norbert Werner} 
\author[\space]{the HERMES-SP collaboration}

\affil[a]{Institute of Physics, E\"otv\"os Lor\'and University, Budapest, Hungary}
\affil[b]{Wigner Research Centre, Budapest, Hungary}
\affil[c]{Department of Theoretical Physics and Astrophysics, Faculty of Science, Masaryk University, Brno, Czech Rep.}
\affil[d]{Astronomical Institute of Charles University, Prague, Czech Rep.}
\affil[e]{University of Udine, Udine, Italy}
\affil[f]{INAF/Astronomical Observatory of Trieste,  Trieste, Italy}
\affil[g]{INAF/Astrophysical and Space Science Observatory (OAS), Bologna, Italy}
\affil[h]{INFN-Sezione di Bologna, Bologna, Italy}

\authorinfo{Further author information: (Send correspondence to G.~Galg\'oczi)\\ G.~Galg\'oczi: E-mail: \url{galgoczi@caesar.elte.hu}, Telephone: +36 30 497 84 25}

\begin{document} 
\maketitle

\begin{abstract}
A software toolkit for the simulation of activation background for high energy detectors onboard satellites is presented on behalf of the HERMES-SP \cite{hermes} collaboration. The framework employs direct Monte Carlo and analytical calculations allowing computations two orders of magnitude faster and more precise than a direct Monte Carlo simulation. The framework was developed in a way that the model of the satellite can be replaced easily. Therefore the framework can be used for different satellite missions. As an example, the proton induced activation background of the HERMES CubeSat is quantified. 
\end{abstract}


\section{INTRODUCTION}
\label{sec:intro}  

The High Energy Rapid Modular Ensemble of Satellites (HERMES) mission \cite{hermes} consists of a fleet of 3U CubeSats hosting broadband X and gamma-ray detectors \cite{evangelista}. The aim of the mission is to detect transient sources with high sky coverage and good angular resolution. The CubeSats will utilize triangulation to locate the investigated X-ray sources. Therefore proton induced activation of these satellites have to be taken into account since it increases background in the long term significantly. An important factor to evaluate for the HERMES mission is the activation background produced by the satellite passages inside the regions with high fluxes of trapped protons, such as the South Atlantic Anomaly (SAA). Different orbits will have different characteristics (depth and duration) and therefore different irradiation doses. Moreover, decay times of all the radio-activated isotopes shall be evaluated. We have developed a framework similar to Odaka et al. \cite{hitomi} which is much more efficient from a computational point of view with respect to a full-scale Monte Carlo simulation. This methods consists of three steps.
\begin{enumerate}
\item A simulation to determine the production rate of each long lived unstable nuclear isotope.
\item Analytical calculation of the decay rates and chains of each isotope at any given time.
\item A second simulation involving only the decaying isotopes and their products.
\end{enumerate}

\section{Determining the number of primary nuclear isotopes}

\begin{figure}[h!]
\centering 
\includegraphics[width=.6\textwidth,origin=c]{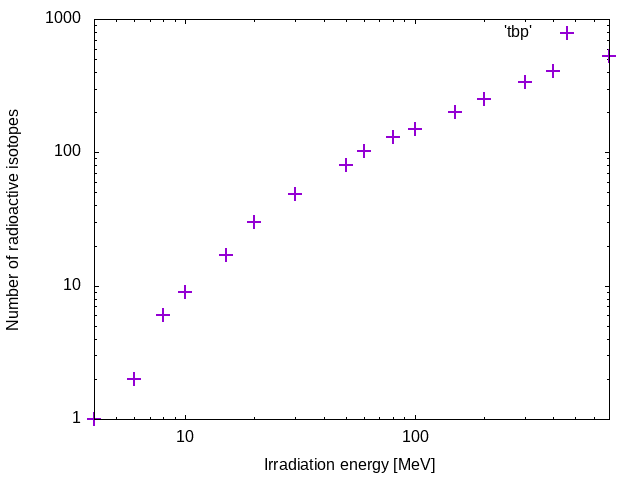}
\caption{\label{fig:noofrad} Number of identified species of isotopes created by proton irradiation on the HERMES satellite for each irradiation energy. Different volumes implemented in the mass model are treated independently, but for this plot they are summed together.}
\end{figure}

\begin{figure}[h!]
\centering 
\includegraphics[width=.6\textwidth,origin=c]{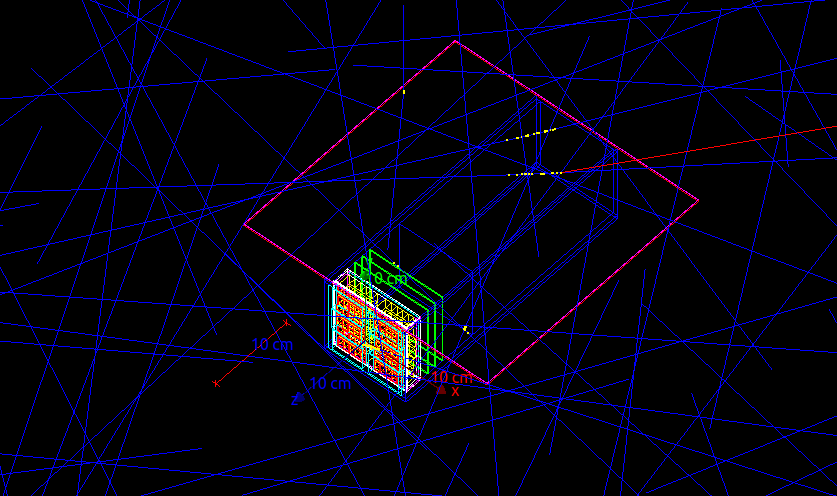}
\caption{\label{fig:second} The mass model of the HERMES satellite being irradiated by protons.}
\end{figure}

The framework which is presented in this paper is based on Geant4 \cite{geant4}.
In the first step a mass model is used to simulate an irradiation of the satellite with monochromatic proton fluxes. One can think of this step as the determination of response to the proton irradiation similarly to the Green function technique. 16 energies between 4 MeV and 700 MeV were simulated. Energies lower than 4 MeV does not create activation in our case and energies above 700 MeV have typically very low fluxes. In Figure \ref{fig:noofrad} the number of different identified isotopes is shown, for a simulation with the HERMES mass model and in Figure \ref{fig:second} the simulation is shown. Higher proton energy results in more possible reactions.

In the “SteppingAction” Geant4 class radioactive decays are detected. Then to prevent recording a decay chain several times, the decayed isotope and the simulation of its secondary particles is stopped. For each volume of the geometry the name, energy, and excitation level of these isotopes are stored with their excitation energy.

Each of the “Physical” volumes in the mass model have to be treated independently. The reason for this is due to two factors. Each volume consists of different materials, therefore different radioactive isotopes will be created. The other reason is that the chance that a decaying isotope inside different volumes will be actually detected can be very different: e.g., an isotope decaying in the detector will be certainly detected, but one in the solar panels is very unlikely to be recorded.

Different energy irradiations yield in the creation of very different isotopes as nuclear cross sections vary significantly by energy. Also, the abundance of the created nuclear isotopes is very different. For example, simulating a proton fluence of 100 000 cm$^{-2}$ with energy of 30 MeV yields mostly to the production of Tb155 and Ge68. For this fluence, about 5000 isotopes of each type are created. For an irradiation at 700 MeV, on the other hand, the most abundant produced isotope is O15, with a number of 170 000 generated nuclei. The second most abundant isotope in this case is Ga68 with 82 000 nuclei. Also, at 700 MeV the number different created isotopes  is 20 times higher than at 30 MeV.

\section{Calculation of activity for time of interest}

The input spectrum used was determined by the AP9 model of SPENVIS Monte Carlo mode with 90\% confidence level for an orbit with height of 600 km and inclination of 40$^\circ$. It should be emphasized that this represents, both for the orbit type and for radiation model used, an \emph{absolute worst} case for the HERMES scenario. 
In this case, most of the protons reach the satellite when it is passing through the SAA. In order to obtain a mean SAA flux, the proton spectrum was averaged simulating an orbit through 30 days sampled at a step of 10 s. Then the fluence of the protons inside one single pass through SAA was determined by multiplying the average spectrum by the transit duration 60$\times$90 s (there is one SAA transit each 90 minutes; therefore the fluence in a single passing corresponds to 60$\times$90 s of the average spectrum assuming that the proton fluence is zero outside of SAA). In this case, an integral proton fluence of 2.5$\times$10$^{6}$ protons/cm$^2$  at energies above 4 MeV was reached. 

\begin{figure}[h!]
\centering 
\includegraphics[width=.6\textwidth,origin=c]{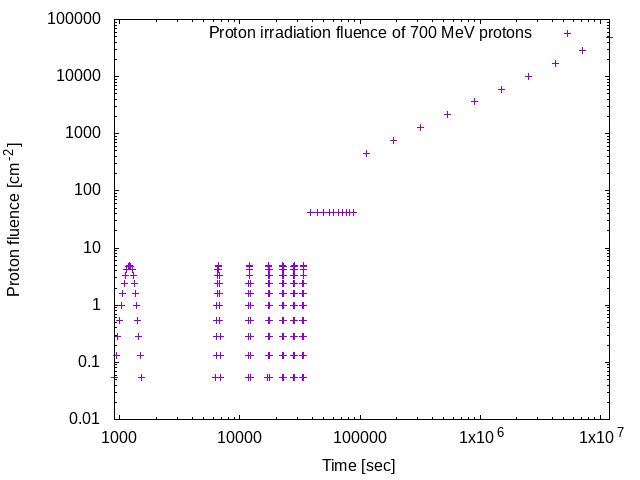}
\caption{\label{fig:irradwindow} Irradiation profile used to calculate activity for 700 MeV protons.}
\end{figure}

The contribution of each SAA passage is different since the created radioactive isotopes have different time elapsed to decaying. Also each isotope has a different decay time, therefore the contribution of each SAA will vary by time. We can solve the Bateman equation\cite{bateman} for each 1 second timestep inside each SAA transit. Therefore we choose to group the decaying times into "time slices". The seven SAA passages closest to the time of interest were sampled in 21 points. The following 9 SAA transits were treated as a single irradiation. The rest of the passages were grouped into 10 irradiation with a logarithmical integration window. In Figure \ref{fig:irradwindow} the described irradiation profile can be seen.

The first seven SAA transits were simulated with a Gaussian time profile sampled in 21 points. The standard deviation of the Gaussian was chosen to be 100 seconds. In order to keep computation time reasonable but still keep precision the 9 passes were assumed to be single irradiation. The rest of the passes were divided into 9 irradiations in time. As the decays have a logarithmic nature, these 9 irradiations were grouped logarithmically in time as can be seen in Figure \ref{fig:second}. 

\begin{figure}[h!]
\centering 
\includegraphics[width=.6\textwidth,origin=c]{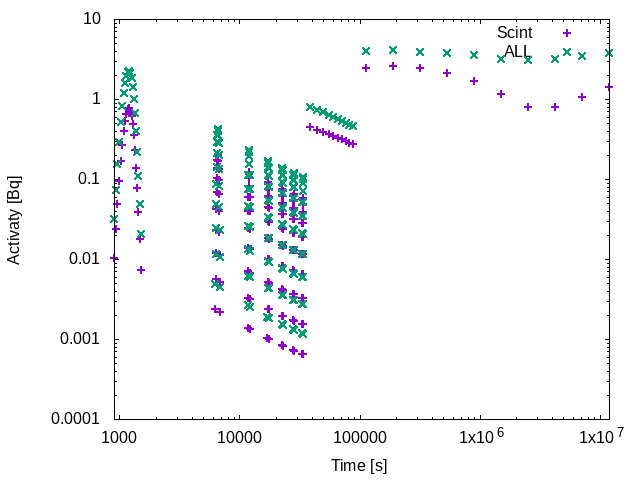}
\caption{\label{fig:no} The activation for each irradiation timeslice from 100 MeV protons for all of the volumes and for only the scintillator. Roughly half of the total satellite activation is due the scintillator.}
\end{figure}

The X-axis of Figure \ref{fig:no} shows time between the time of interest and the given irradiation. For each of these points the decays chains are calculated and then summed up to yield the total number of isotopes.  The explicit formula to determine the amount of the N$^{th}$ isotope in a decay chain after and elapsed time $t$ time is:

\begin{equation}
N_{n} (t) = \sum_{i=1}^{n}  \left[ N_i(0) \times \left(  \prod_{j=i}^{n-1} \lambda_j \right) \times  \left( \sum_{j=i}^{n} \left( \frac{e^{-\lambda_j t}}{ \prod_{p=i,p \neq j}{n} ( \lambda_p - \lambda_j } \right) \right) \right] 
\end{equation}

The algorithm identifies the decaying isotopes created by the proton-induced activation in the first simulation. Most of them are in ground state. Afterwards the algorithm identifies the decay chains and the decay rates of the corresponding daughter nuclei before the deletion of very short lived particles. 

The first step in this algorithm is the construction of the decay chains, which describe the decay of the isotopes identified in the first simulation. This is implemented in C++, employing the built-in function of Geant4 to access the decay products of the element in question. The decay tables are taken from the ESNDF dataset included in the Geant4 distribution. Excited states and floating levels (same excitation energy but different angular momentum) of the nuclei are treated individually. (An example for the importance of this: for Ta178, half life can be an order of magnitude different in case of different floating level of the nuclei.) The decay chains were created by identifying the decay modes of the original isotopes and deducing what the daughter isotope(s) can be.  The implemented algorithm  looks recursively for the unstable daughters of the isotope in question. This was repeated until the last element has no unstable daughters. In order to keep the number of decay chains reasonable, daughter isotopes with branching ratios lower than a given threshold are neglected.

An example of such a decay chain is:

Hf156  \textrightarrow Yb152  \textrightarrow Tm152[482.320]  \textrightarrow Tm152  \textrightarrow Er152[1715.400]  \textrightarrow Er152  \textrightarrow Ho152[179.400]  \textrightarrow Ho152  \\
\textrightarrow Dy152[3500.000]  \textrightarrow Dy152 
\textrightarrow Tb152[256.930]  \textrightarrow Tb152  \textrightarrow Gd152[2880.670]  \textrightarrow Gd152  \textrightarrow Sm148  \textrightarrow Nd144.

In the brackets is shown the excitation energy in keV. The number of decay chains increases by input proton energy, all together there are more than 1.5 million chains. The Bateman equation \cite{bateman} has to be solved for each of these chains for every time in the irradiation profile. Moreover, the decay chains are treated independently in each volume. Still on a standard PC (Intel Core i7-8700K CPU @ 3.70GHz × 12, 32 GB of RAM) the algorithm runs in a few hours by setting a lower threshold for the cutoff of branching ratios to 5\%.

The second part of the algorithm is the exclusion of isotopes (mostly isomers) with very short half life from the decay chains. This is needed since to obtain the amount of each isotope after a given time, one needs to solve the Bateman equation. In this equation, for very large decay rates ($\lambda$) even the double-precision floating-point format is not sufficient. Therefore the isotopes with a decay rate larger than $10^9$ are erased from the decay chains: this corresponds to a half life of 21 ms. This includes only very short lived isomers (like Ta181[6.237] or W181[365.550]) which would not be abundant after the time in question passes. Therefore it does not affect the results of the simulation.  

\begin{figure}[h!]
\centering 
\includegraphics[width=.6\textwidth,origin=c]{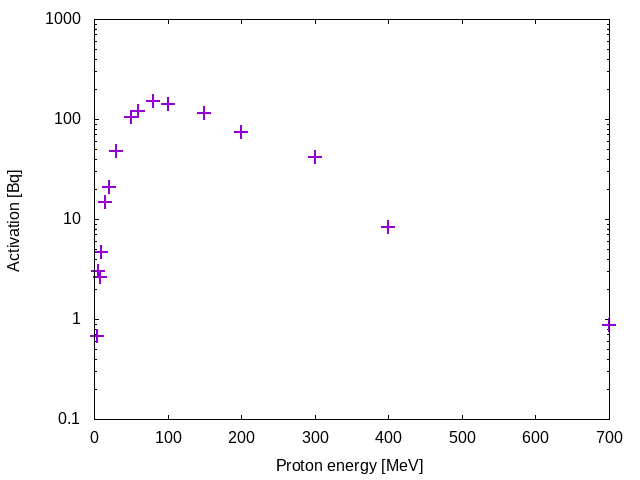}
\caption{\label{fig:contribut} The contribution of different energy bands of primary protons to the activation of the satellite.}
\end{figure}

The third part of this step is to solve the Bateman equation for these decay chains. The solution of the Bateman equation does not have a closed form, although it was possible to implement the solution recursively in C++. In order to be able to solve the equation, the decay chains were created in linear fashion. In order to obtain the activity of each radioactive isotope, their amount needs to be determined. The software determines this normalization factor from the provided integral photon flux.

\begin{figure}[h!]
\centering 
\includegraphics[width=.6\textwidth,origin=c]{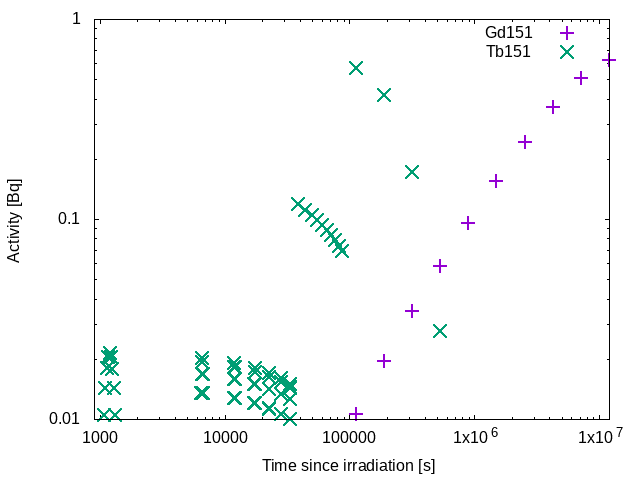}
\caption{\label{fig:design2} The activity of Tb151 and Gd151 in different timeslices with respect to the time of interest. The previous decays into the latter. For long timescales the latter is more important for shorter ones, the previous.}
\end{figure}

As can be seen in Figure \ref{fig:contribut} about half of the activation due to 100 MeV protons is created in the Gadolinium Aluminium Gallium Garnet (GAGG) scintillator. For other energies this is different as different isotopes have different energy dependence on proton activation cross section. Also what can be seen is that the activity of the volumes changes by time as the isotopes in the scintillator decay "faster" than those on the other satellite structures. The proton energy that contributes most to the overall activity in the HERMES case is about 100 MeV, since protons around this energy have intermediate flux and also quite high activation cross-section compared to lower energies.

For different timescales different isotopes dominate the activity. For example let's take the decay chain: 

Tb151 \textrightarrow Gd151[839.320] \textrightarrow  Gd151 \textrightarrow  Eu151 \textrightarrow  Pm147 \textrightarrow  Sm147

Tb151 decays with a half-life of 18 hours. After about a day Gd151 becomes more important than Tb151 as can be seen in Figure \ref{fig:design2}.
For the simulated orbit, the activity of the whole satellite is 851 Bq after half a year in space assuming the discussed worst-case irradiation profile (AP9 90\% c.l. model, 600 km altitude, inclination of 40$^\circ$).

\section{Simulation of detector response}

\begin{figure}[h!]
\centering 
\includegraphics[width=.6\textwidth,origin=c]{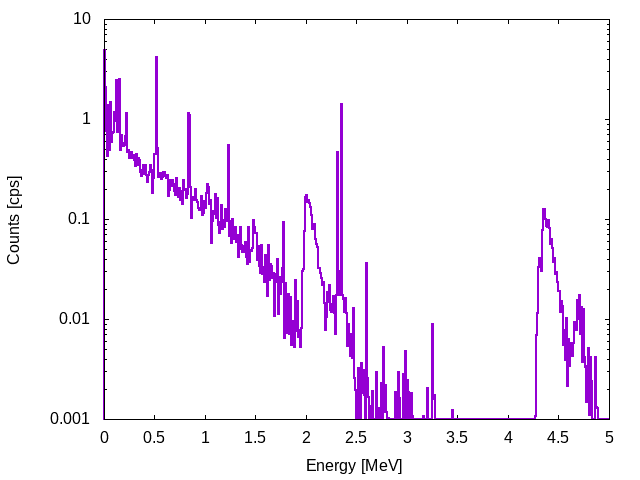}
\caption{\label{fig:enspec} Energy spectrum deposited inside the GAGG scintillator by the proton induced nuclear decays after half a year in space.}
\end{figure}

The final result of the last algorithm is a list of isotopes sorted by activity in each volume of the mass model. In this simulation the most active isotopes are put in the respective volume where they were created and their decay is simulated. The energy deposited inside the Silicon Drift Detector (SDD) and the scintillator is registered. The histogram of the energy depositions are normalised to the isotope in question. In the end these histograms are merged into one single histogram which represents what we would actually measure in space due to activation (assuming a perfect energy resolution). Figure $\ref{fig:enspec}$ shows the merged histogram obtained when assuming half a year spent in space.

The deposited energy inside the scintillator can be seen in Figure \ref{fig:enspec}. One can notice the $\beta$+ decay peak at 511 keV and also the characteristic electron capture and $\gamma$ peaks. \textbf{Assuming a lower detector threshold of 20 keV the measured background would be 60 counts per second after spending half a year in space.} It is interesting to compare this to the 800 Bq activity of the satellite. The deposited energy inside the SDD can be seen in Figure \ref{fig:sdd}. The number of detections originating in the SDD is negligible compared to the ones in the scintillator as the SDD is much thinner and consists of material with lower atomic number.

\begin{figure}[h!]
\centering 
\includegraphics[width=.5\textwidth,origin=c]{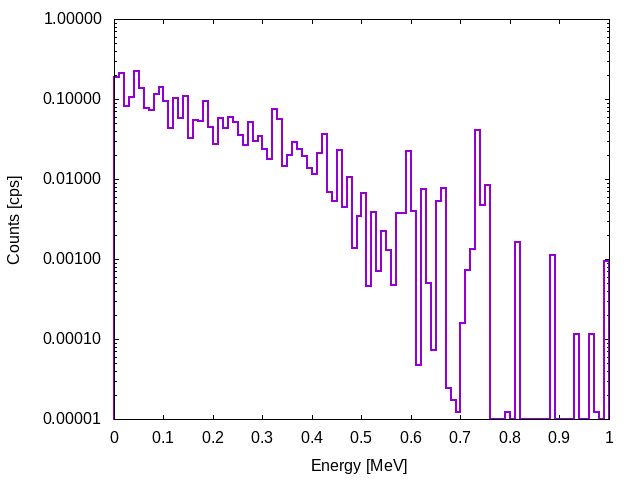}
\caption{\label{fig:sdd} Deposited energy spectrum of the SDD by nuclear decays from activation.}
\end{figure}

\section{Validation of the framework}

In order to ensure the validity of the results of the implemented simulation framework both direct Geant4 simulations and activation measurements from literature were used. In the direct simulations, Geant4 did all the calculations regarding the decaying nuclei. Energy depositions between 10692 minutes and 10908 minutes after the irradiation were recorded in this case to be compared to the results of our framework and in the experimental results of Sakano et al. \cite{sakano}. A GAGG crystal with dimensions of $5.0 \times 5.0 \times 40$ mm$^3$ was irradiated by a proton beam with an energy of 150 MeV. A photomultiplier tube was used to read out the activation signal.

\begin{figure}[h!]
\centering 
\includegraphics[width=.55\textwidth,origin=c]{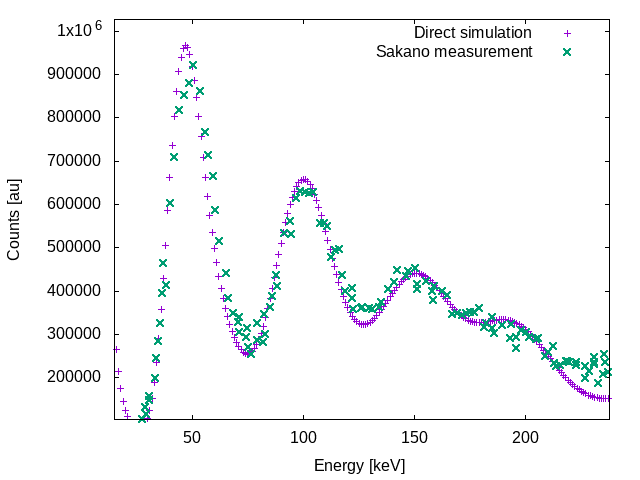}
\caption{\label{fig:direct_vs_sakano} Deposited energy in a single GAGG crystal after 10800 minutes of irradiation by protons with an energy of 150 MeV. The direct simulation performed by us matches the experimental results of Sakano et al. \cite{sakano}.}
\end{figure}

To be able to run the direct simulations in a reasonable timeframe, only a single GAGG crystal was irradiated by 10$^{9}$ protons with an energy of 150 MeV. In Figure \ref{fig:direct_vs_sakano} the energy spectrum obtained in direct Geant4 simulations due to proton induced activation is compared to measurements obtained by Sakano et al. The results of the simulation were convolved by a Gaussian function with changing width to include the statistical process of photon detection in the measurements. Since the used measurements only span the low energy regime (up to 200 keV) we decided to validate our framework with direct Geant4 simulations. 

\begin{figure}[h!]
\centering 
\includegraphics[width=.6\textwidth,origin=c]{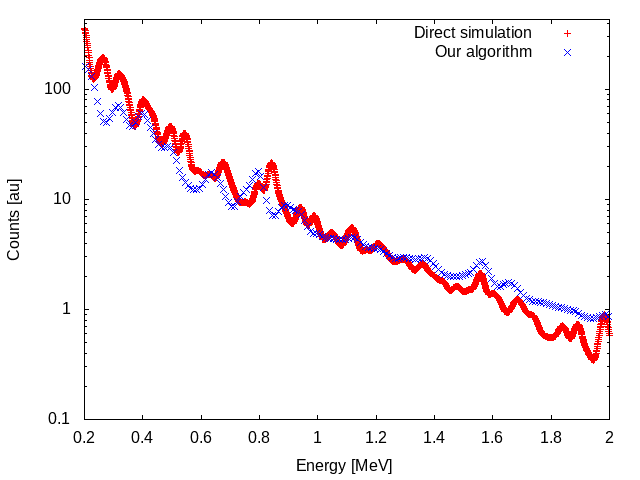}
\caption{\label{fig:direct_vs_custom} Deposited energy in a single GAGG crystal after 10800 minutes of irradiation by protons with an energy of 150 MeV. The direct simulation performed by us matches the results of our custom framework that we developed.}
\end{figure}

\section{Discussion}

Within the HERMES collaboration a Geant4 based framework was developed to understand and quantify the effect of proton induced activation. The framework utilizes direct Monte Carlo and analytical calculations to be able to run two orders of magnitude faster and more accurate simulation, with respect to a direct Monte Carlo approach. The framework was developed in a way that the geometry of the satellite can be replaced easily. We have applied the framework to our satellite and determined energy deposition spectrum in the HERMES scintillators and in the SDD detectors as well after half a year in space. Assuming a low energy threshold of 20~keV the background due to proton activation would be 60 cps which is within the required specifications \cite{campana}.


\clearpage

\appendix    

\section{The construction of the decay chains}
\label{sec:misc}

\begin{figure}[h!]
\centering 
\includegraphics[width=0.99\textwidth,origin=c]{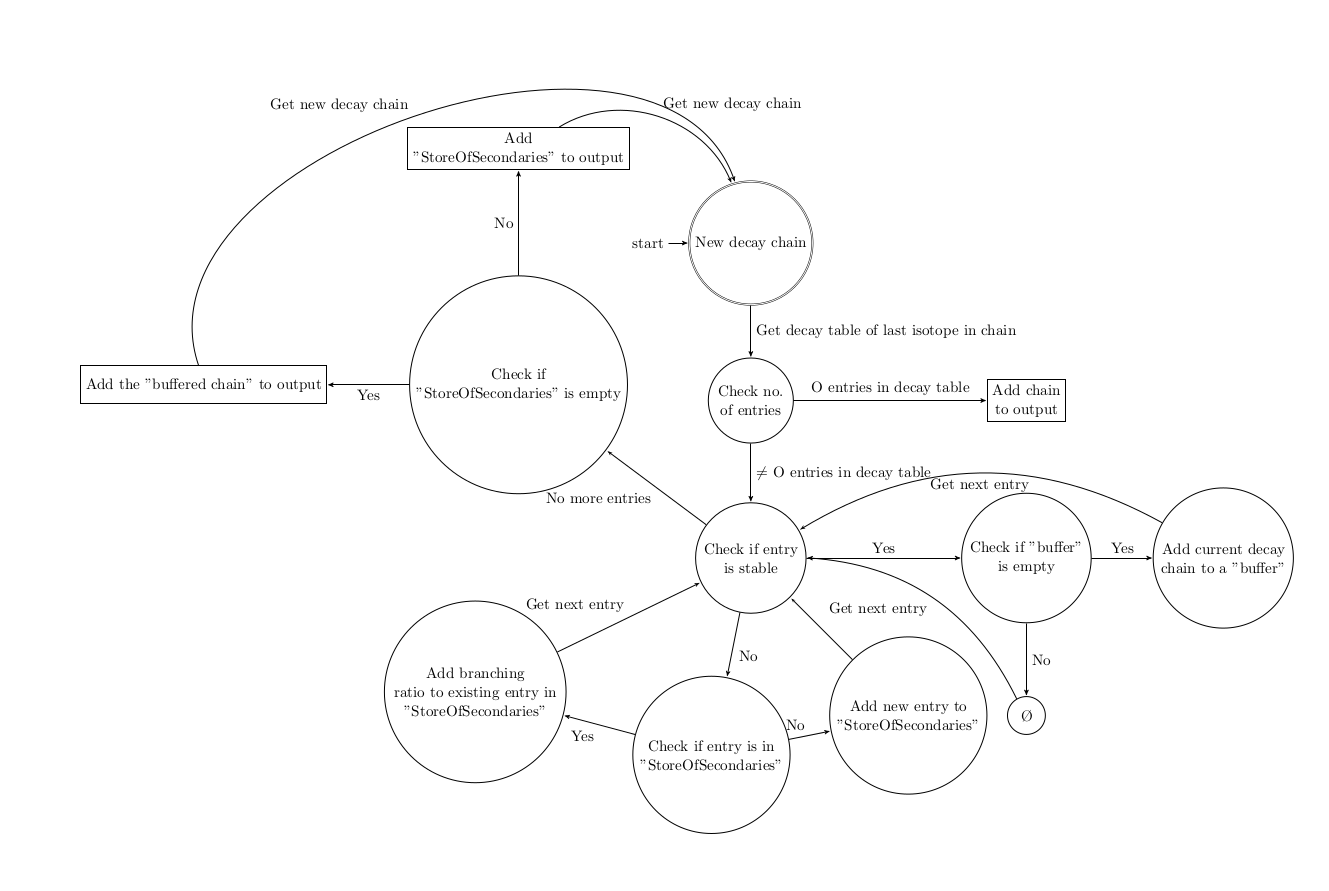}
\caption{\label{fig:sdd} The state diagram chart of the decay chain construction from the decay scheme of the mother nuclei.}
\end{figure}

\acknowledgments 
This project has received funding from the European Union Horizon 2020 Research and Innovation Framework Programme under grant agreement HERMES-Scientific Pathfinder n. 821896 and from ASI-INAF Accordo Attuativo HERMES Technologic Pathfinder n. 2018-10-HH.0.
The research has been supported by the European Union, co-financed by the European Social Fund (Research and development activities at the E\"{o}tv\"{o}s Lor\'{a}nd University's Campus in Szombathely, EFOP-3.6.1-16-2016-00023).
\bibliography{report} 
\bibliographystyle{spiebib} 

\end{document}